\documentstyle[titlepage]{article}
\title{Magnetism of superconducting $U\!Pt_{3}$}
\author{
G Hara\'n\thanks{Supported by the European Community's Action
for Cooperation in Sciences and Technology with Central and Eastern
European Countries Fellowship,
\mbox{contract $n^{\circ}:\; ERB-CIPA-92-2218$.}}
\thanks{permanent address:
Institute of Physics,
Politechnika Wroc{\l}awska,
\mbox{Wybrze\.ze Wyspia\'nskiego 27,}
\mbox{50-370 Wroc{\l}aw, Poland}} and G A Gehring\\
\vspace{0.4cm}\\
{\it Department of Physics}\\
{\it The University of Sheffield}\\
{\it Sheffield S3 7RH}\\
{\it UK} }
\begin{document}
\maketitle

\begin{abstract}
\hspace{0.2cm}The phase diagram of superconducting $U\!Pt_{3}$ in
\mbox{pressure-temperature} plane, together with the neutron scattering
data is studied within a two component superconducting order parameter
scenario. In order to give a qualitative explanation to the experimental
data a set of two linearly independent antiferromagnetic moments which
emerge appropriately at the temperature \mbox{$T_{N}\sim 10\cdot T_{c}$}
and \mbox{$T_{m}\sim T_{c}$} and couple to superconductivity is proposed.
Several constraints on the fourth order coefficients in the
\mbox{Ginzburg-Landau} free energy are obtained.
\end{abstract}

\section{\bf Introduction}
\hspace{0.5cm} A heavy fermion superconducting $U\!Pt_{3}$ compound is an
example of the unconventional superconductivity, in which both the gauge and
the point group symmetries are broken in the ordered phase. At the temperature
$T_{N}\simeq 5K$ it undergoes the antiferromagnetic transition with the
magnetic moments confined to the $D_{6h}$ basal plane, however the long range
antiferromagnetic correlations have not yet been seen \cite{1,2}. Far below
the N\'eel temperature, at $T_{c_{+}}\simeq 0.51K$ (p=0 bar) $U\!Pt_{3}$
becomes superconducting \cite{3,4}. There is another superconducting
transition at $T_{c_{-}}\simeq 0.46K$ (p=0 bar) \cite{3,4}. This feature and
a rich phase diagram in the magnetic field and temperature plane \cite{5} are
accepted as the evidence of a multicomponent superconducting order parameter.
There are also the pressure experiments which indicate strongly
the coupling between superconductivity and magnetism in $U\!Pt_{3}$
\cite{2,7,8}. Namely the specific heat measurements under pressure show that
the two critical temperatures $T_{c_{+}}$ and $T_{c_{-}}$ converge into one
critical temperature $T_{c}$ above $p_{c}\simeq 4$ kbar pressure value Fig.1
\cite{2,7,8}, which is the pressure that destroyes antiferromagnetism in the
system. This experiment supports the theory of a two component order parameter
$\bar{\psi}=(\psi_{x},\psi_{y})$ in a basal plane of the crystal, belonging
to a two-dimensional irreducible representation of the hexagonal point
group $D_{6h}$. In this approach a complex vector $\bar{\psi}$ couples to
the magnetic moment $\bar{M}$ and the split transition is due to that
interaction. The role of magnetism as a symmetry breaking field coupling
to superconductivity is revealed in the neutron scattering measurements
\cite{1,10}. In these experiments Aeppli et al. established that below the
temperature of the order of a superconducting transition temperature the
neutron scattering intensity of the $(1,\frac{1}{2},0)$ reflection suddenly
saturates and is almost constant unless the superconductivity occurs.
There is a remarkable change in the temperature dependence for a
superconducting system. At a temperature of the order of $T_{c}$ the slope
of the neutron scattering intensity changes sign and the intensity becomes
an increasing function of temperature see Fig.2 \cite{1,10}. This is another
strong evidence of the coupling between magnetism and superconductivity
in $U\!Pt_{3}$.\\
\hspace*{0.5cm} Recently Joynt \cite{11} discussed within a two component order
parameter approach the phase diagram of $U\!Pt_{3}$ in three-dimensional
magnetic field-pressure-temperature space. It agrees qualitatively with
measurements \cite{2,5,7,8,12}. However the temperature dependence of the
magnetic moment observed by Aeppli et al. \cite{1,10} was not taken into
account. The contradiction here arises as follows. The magnetic Bragg peak
observed in neutron scattering \cite{1,10} reproduced in Fig.2
shows that the superconductivity is acting to suppress the magnetism.
By thermodynamic reasoning we know that if the onset of superconductivity
reduces the magnetism then the onset of magnetism must reduce the tendency
to superconductivity. The magnetism may be removed by pressure \cite{2,7,8}.
We observe that as the pressure is reduced below $p_{c}$ where the magnetism
reappears the slope of transition temperature is increased see Fig.1.
In other words the critical temperatures $T_{c_{+}}$ and $T_{c_{-}}$ are not
suppressed equally by the pressure, what can be expressed quantitatively by
an inequality which follows :
\begin{eqnarray}
\label{e0}
\frac{T_{c_{+}}\left(p=0\right)-T_{c}\left(p=0\right)}
{T_{c}\left(p=0\right)-T_{c_{-}}\left(p=0\right)} & > & 1\;,
\end{eqnarray}
see Fig.1.\\
We show that this competition between superconductivity and antiferromagnetism
cannot be understood within the simple model of magnetism considered so far.\\
\hspace*{0.5cm} The plan of the paper is as follows. In section 2 we study
mentioned already pressure and magnetic experiments in a frame of
\mbox{two-dimensional} superconducting order parameter scenario. To avoid
the inconsistencies following from this approach we introduce  a two magnetic
moment model in section 3. Within this scenario we analyse the experimental
data and obtain several constraints on the \mbox{Ginzburg-Landau} free energy
coefficients in sections 3 and 4. Finally we summarize the results in section
5.

\section{\bf Two component superconductivity coupled to magnetism}
\hspace{0.5cm} In this section we review the experimental evidence which
supports this model and then construct the free energy. The free energy
is used to obtain the coupled order parameters of magnetism and
superconductivity. This analysis follows \cite{9,16,17}. We reproduce it here
because it is important to consider both the temperature and pressure
experiments using a unified notation. In this approach we start with a free
energy density :
\begin{equation}
\label{e1}
F=F_{M}+F_{S}+F_{SM},
\end{equation}
where
\begin{equation}
\label{e2}
F_{M}=\left\{
\begin{array}{ccl}
\alpha_{M}(T^{\ast}-T_{N})M^{2}+\frac{1}{2}\beta_{M}M^{4}, & for & T
\leq T^{\ast}\\
\alpha_{M}(T-T_{N})M^{2}+\frac{1}{2}\beta_{M}M^{4}, & for & T > T^{\ast}
\end{array}
\right.
\end{equation}
\begin{equation}
\label{e3}
F_{S}=\alpha_{S}(T-T_{c})|\bar{\psi}|^{2}+\frac{1}{2}\beta_{1}|\bar{\psi}|^{4}+
\frac{1}{2}\beta_{2}|\bar{\psi}^{2}|^{2},
\end{equation}
\begin{equation}
\label{e4}
F_{SM}=\gamma|\bar{M}\bar{\psi}|^{2}+\alpha M^{2}|\bar{\psi}|^{2}.
\end{equation}
All the Ginzburg-Landau coefficients are very weakly temperature and pressure
dependent, what can be shown in a weak-coupling microscopic theory \cite{14},
hence we choose them to be constant. The magnetic free energy given by
equation $\mbox{(\ref{e2})}$ has been chosen to include the phenomenological
saturation of $M$ below $T^{\ast}$ \cite{1,10}.
The coefficients in $F_{M}\;
\mbox{(\ref{e2})}$ and $F_{S}\;\mbox{(\ref{e3})}$ are positive whereas the
$\gamma$ coefficient in $F_{SM}\;\mbox{(\ref{e4})}$
may be chosen to be negative and $\bar{M}$ is then parallel to $\hat{x}$.
The superconducting order parameter $\bar{\psi}=(\psi_{x},\psi_{y})$ is
complex and its composits $\psi_{x}$ and $\psi_{y}$ are written as
\mbox{$\psi_{x}=|\psi_{x}|{\mit e}^{{\mit i}\varphi_{x}}$} and
\mbox{$\psi_{y}=|\psi_{y}|{\mit e}^{{\mit i}\varphi_{y}}$.}\\
\hspace*{0.5cm}Minimisation of  the free energy leads to the following
equations for the order parameters :
\begin{equation}
\label{e01}
0=\alpha_{M}(T_{M}-T_{N})+\beta_{M}M^{2}+\gamma|\psi_{x}|^{2}+
\alpha|\bar{\psi}|^{2}\;\;\;{\rm or}\;\;\;M=0\;,
\end{equation}
where
\begin{displaymath}
T_{M}=\left\{
\begin{array}{ccl}
T\;, & for & T>T^{\ast}\\
T^{\ast}\;, & for & T\leq T^{\ast}\;,
\end{array}
\right.
\end{displaymath}
\begin{equation}
\label{e02}
0=\alpha_{S}(T-T_{c})+\beta_{1}|\bar{\psi}|^{2}+\beta_{2}(|\psi_{x}|^{2}+
|\psi_{y}|^{2}\cos 2(\varphi_{x}-\varphi_{y}))+\gamma M^{2} +
\alpha M^{2}
\end{equation}
\hspace*{0.3cm} or $\;\;\;|\psi_{x}|=0$,
\begin{equation}
\label{e03}
\left\{
\begin{array}{l}
0=\alpha_{S}(T-T_{c})+\beta_{1}|\bar{\psi}|^{2}+\beta_{2}(|\psi_{y}|^{2}+
|\psi_{x}|^{2}\cos 2(\varphi_{x}-\varphi_{y}))+\alpha M^{2}\\
{\rm and}\;\;\varphi_{x}-\varphi_{y}=\frac{\pi}{2}
\end{array}
\right.
\end{equation}
\hspace*{0.5cm} or $\;\;\;|\psi_{y}|=0$.\\
\vspace{0.3cm}\\
\hspace*{0.5cm} From these expressions we find the following conditions for
$M$, $\psi_{x}$ and $\psi_{y}$ :
\begin{equation}
\label{e04}
M=|\psi_{x}|=|\psi_{y}|=0\;\;\;\;for\;\;\;\;T>T_{N}\;,
\end{equation}
\begin{equation}
\label{e05}
M^{2}=\frac{\alpha_{M}}{\beta_{M}}(T_{N}-T),\;\;\;\;|\psi_{x}|=|\psi_{y}|=0
\;\;\;\;for\;\;\;\;T^{\ast}<T \leq T_{N}\;,
\end{equation}
\begin{equation}
\label{e06}
M^{2}=\frac{\alpha_{M}}{\beta_{M}}(T_{N}-T^{\ast}),\;\;\;\;|\psi_{x}|=
|\psi_{y}|=0\;\;\;\;for\;\;\;\;T_{c_{+}}<T\leq T^{\ast}\;,
\end{equation}
\begin{equation}
\label{e07}
\left\{
\begin{array}{l}
M^{2}=\frac{1}{\beta_{M}}\left[\alpha_{M}(T_{N}-T^{\ast})-(\gamma+\alpha)
|\psi_{x}|^{2}\right]\\[0.2cm]
|\psi_{x}|^{2}=\frac{\alpha_{S}}{\beta_{1}+\beta_{2}}(T_{c_{+}}-T)\\[0.2cm]
|\psi_{y}|=0
\end{array}
\right.\;\;for\;\;T_{c_{-}}<T\leq T_{c_{+}}\;,
\end{equation}
\begin{equation}
\label{e08}
\left\{
\begin{array}{l}
M^{2}=\frac{1}{\beta_{M}}\left[\alpha_{M}(T_{N}-T^{\ast})-(\gamma+\alpha)
|\psi_{x}|^{2}-\alpha|\psi_{y}|^{2}\right]\\
|\psi_{x}|^{2}=\frac{1}{\beta_{1}+\beta_{2}}\left[\alpha_{S}(T_{c_{+}}-T)
-(\beta_{1}-\beta_{2})|\psi_{y}|^{2}\right]\\
|\psi_{y}|^{2}=\frac{\alpha_{S}}{\beta_{1}+\beta_{2}}(T_{c_{-}}-T)
\end{array}
\right.\;\;for\;\;T\leq T_{c_{-}}\;,
\end{equation}
where
\begin{equation}
\label{e09}
T_{c_{+}}=T_{c}-\frac{\gamma+\alpha}{\alpha_{S}}M^{2}\;,
\end{equation}
\begin{equation}
\label{e00}
T_{c_{-}}=T_{c}-\frac{1}{\alpha_{S}}\left[\alpha M^{2}+(\beta_{1}-\beta_{2})
|\psi_{x}|^{2}\right]\;.
\end{equation}
$T_{c}$ is the superconducting transition temperature in a system without the
magnetism. The complete solution to \mbox{Eqs.$(\ref{e04})-(\ref{e08})$} that
is the explicit formulae for $T_{c_{-}}$ and $T_{c_{+}}$ are given in
Appendix A \mbox{$(\ref{a01})-(\ref{a02})$}. The magnetic moment changes as
\mbox{$M^{2}=\frac{\alpha_{M}}{\beta_{M}}(T_{N}-T)$} for temperatures higher
than temperature $T^{\ast}$, then suddenly saturates at \mbox{$T^{\ast}\;\;
(T^{\ast} \sim T_{c})$} and becomes constant below this temperature :
\mbox{$M^{2}=\frac{\alpha_{M}}{\beta_{M}}(T_{N}-T^{\ast})$} in a normal
(not superconducting) state.
This temperature dependence of the magnetic moment is consistent with the
measurements by Aeppli et al \cite{1,10}. They observed a kink at
\mbox{$T^{\ast} \sim T_{c}$} and almost constant value of the magnetic
Bragg intensity below \mbox{$T^{\ast}$} for magnetic field
\mbox{$H>H_{c_{2}}$} that is when the system was not superconducting.
The \mbox{$T^{\ast}$} temperature is introduced in our free energy
\mbox{$(\ref{e2})$} rather artificially in order to fit the existing
experimental data \cite{1,10}. We shall comment more on this issue further
in the text.\\
\hspace*{0.5cm} From the free energy density $F_{S}\;\;(\ref{e3})$ we get the
linear pressure dependence of the superconducting transition temperature:
\begin{equation}
\label{e5}
T_{c}=T_{c}^{0}-a_{0}p\;,
\end{equation}
where $a_{0}$ is a constant coefficient and $T_{c}^{0}$ - a critical
temperature $T_{c}$ at zero pressure $(p=0)$. We also assume the squared
magnetic moment to be a linear pressure function:
\begin{equation}
\label{e6}
M^{2}=M_{0}^{2}\frac{p_{N}-p}{p_{N}}\;,
\end{equation}
where $M_{0}$ is a magnetic moment at $p=0$, \mbox{$M_{0}=M(T,p=0)$}
and $p_{N}$ $(\mbox{$p_{N}$}=\mbox{$p_{c}\simeq 4 kbar$})$ is a pressure at
which the antiferromagnetism vanishes. In the superconducting system described
by the free energy density $(\ref{e1})$ the magnetic and the superconducting
terms compete in the coupling term $(\ref{e4})$. This interaction leads to the
split of critical temperature \mbox{$T_{c}$} into \mbox{$T_{c_{-}}$} and
\mbox{$T_{c_{+}}$} \cite{9} :
\begin{equation}
\label{e7}
T_{c_{+}}-T_{c_{-}}=\frac{|\gamma|}{\alpha_{S}}\frac{\beta_{1}+\beta_{2}}
{2\beta_{2}}M^{2}\;.
\end{equation}
\hspace{0.5cm} One can establish the pressure dependence of \mbox{$T_{c_{+}}$}
and \mbox{$T_{c_{-}}$} from Eqs.$(\ref{e5})$ and $(\ref{e6})$ :
\begin{eqnarray}
\label{e8}
T_{c_{+}} & = & T_{c_{+}}^{0}-a_{+}p\;,\\
\label{e9}
T_{c_{-}} & = & T_{c_{-}}^{0}-a_{-}p\;,
\end{eqnarray}
where
\begin{eqnarray}
\label{e10}
T_{c_{+}}^{0} & = & T_{c}^{0}+\frac{|\gamma|-\alpha}{\alpha_{S}}M_{0}^{2}\;,\\
\label{e11}
T_{c_{-}}^{0} & = & T_{c}^{0}-\frac{1}{\alpha_{S}}\left(\alpha+
\frac{\beta_{1}-\beta_{2}}{2\beta_{2}}|\gamma|\right)M_{0}^{2}\;,\\
\label{e12}
a_{+} & = & a_{0}+\left(\frac{|\gamma|-\alpha}{\alpha_{S}}\right)
\frac{M_{0}^{2}}{p_{N}}\;,\\
\label{e13}
a_{-} & = & a_{0}-\frac{1}{\alpha_{S}}\left(\alpha+\frac{\beta_{1}
-\beta_{2}}{2\beta_{2}}|\gamma|\right)\frac{M_{0}^{2}}{p_{N}}\;.
\end{eqnarray}
\hspace{0.5cm} To obtain a proper pressure behavior (Fig.1) the following
constraints must be fulfilled :
\begin{equation}
\label{e14}
a_{+}>a_{0} \;\;\;\;\;{\rm and}\;\;\;\;\;a_{-}<a_{0}\;.
\end{equation}
Together with a condition $(\ref{e0})$ they give the relations between \\
the \mbox{Ginzburg-Landau} coefficients :
\begin{equation}
\label{e15}
\frac{1}{2}\left(1-\frac{\beta_{1}}{\beta_{2}}\right)|\gamma|<\alpha<
\frac{1}{4}\left(3-\frac{\beta_{1}}{\beta_{2}}\right)|\gamma|\;.
\end{equation}
\hspace{0.5cm} Now we turn to the magnetic Bragg scattering measurements
\cite{1,10} (Fig. 2a). Since the neutron scattering intensity is
proportional to
$M^{2}$ we look at the magnetic moment and analyse it as a function of
temperature. Taking into
account that the coupling coefficients $\alpha$ and $\gamma$ $(\ref{e4})$
are expected to be much smaller than the other G-L coefficients \cite{14}
and therefore neglecting higher than the linear in $\alpha$ and $\gamma$
terms from \mbox{Eqs.$(\ref{e1})-(\ref{e4})$} we obtain~:
\begin{equation}
\label{e16}
M^{2}=M_{c}^{2}+a_{M}T\;,
\end{equation}
where
\begin{equation}
\label{e17}
M_{c}^{2}=\frac{\alpha_{M}}{\beta_{M}}(T_{N}-T^{\ast})-a_{M}T_{c}\;,
\end{equation}
and
\begin{eqnarray}
\label{e18}
a_{M}=\frac{\alpha_{S}}{\beta_{M}}\frac{\gamma+\alpha}{\beta_{1}+\beta_{2}} &
for & T_{c_{-}}<T\leq T_{c_{+}}\;,\\
\label{e19}
a_{M}=\frac{\alpha_{S}}{\beta_{M}}\frac{2\alpha+\gamma}{2\beta_{1}} & for &
T \leq T_{c_{-}}\;.
\end{eqnarray}
In $M^{2}$ given by \mbox{Eqs.$(\ref{e16})-(\ref{e19})$} a discontinuity
arises at \mbox{$T=T_{c_{-}}$} with a jump of the second order of magnitude
in $\alpha$ and $\gamma$. Therefore it is negligible in the linear
approximation.
We present the full formula for $M^{2}$
in Appendix A
\mbox{Eqs.$(\ref{a03})-(\ref{a07})$}. It can be shown that even within this
general description the results of this section still hold.\\
There are two characteristic temperatures - $T_{c_{+}}$ and $T_{c_{-}}$
distinguished by the superconducting phase transitions, hence the change
in the temperature dependence of the magnetic moment due to superconductivity
can take place at one of these temperatures. For $M^{2}$ increasing with
the temperature up to $T_{c_{+}}$ and decreasing then, that is for a kink at
$T=T_{c_{+}}$ the condition :
\begin{eqnarray}
\label{e20}
a_{M}>0 & for & T<T_{c_{+}}
\end{eqnarray}
is required, while for a kink at $T=T_{c_{-}}$ the following constraints are
to be fulfilled:
\begin{eqnarray}
\label{e21}
a_{M}<0 & for & T_{c_{-}}<T<T_{c_{+}}
\end{eqnarray}
and
\begin{eqnarray}
\label{e22}
a_{M}>0 & for & T<T_{c_{-}}\;.
\end{eqnarray}
The condition $(\ref{e20})$ leads to the inequality:
\begin{equation}
\label{e23}
\alpha>|\gamma|\;,
\end{equation}
whereas from $(\ref{e21})\;{\rm and}\;(\ref{e22})$ follows that:
\begin{equation}
\label{e24}
\frac{1}{2}|\gamma|<\alpha<|\gamma|\;.
\end{equation}
It is evident that the condition $(\ref{e23})$ is inconsistent with the
pressure relation $(\ref{e15})$, while the conditions $(\ref{e15})$ and
$(\ref{e24})$ yield the relation \mbox{$\frac{\beta_{1}}{\beta_{2}}<1$}
which is in contradiction with the specific heat measurements data
\cite{15}. Put into words thermodynamics requires that if the magnetic moment
is reduced when the sample becomes superconducting then the tendency to become
superconducting will be increased if the magnetism is removed. This implies
that the continuation of the phase line between normal and superconducting
phases for $p>p_{c}$ should lie above $T_{c_{+}}$ if it is extrapolated
back to low pressure in clear contrast to the data shown in Fig.1 and also
more recent data of Boukhny et al. \cite{20}.\\
\hspace*{0.5cm} Therefore we conclude that it is {\bf not possible to explain
the pressure and neutron scattering data in a frame of the free energy density
${\bf (\ref{e1})-(\ref{e4})}$} and the decrease in the magnetic Bragg intensity
cannot be attributed to the decrease in $M$ only if it is assumed that
$T_{c_{+}}-T_{c_{-}}$ is due to the coupling with magnetism. This paper does
not address the alternative possibility that the splitting of $T_{c}$ is due
instead to the coupling of the superconductivity to the charge density wave
\cite{19,21} except to note that even if the effect of magnetism is only to
reduce both $T_{c_{+}}$ and $T_{c_{-}}$ due to a pair breaking mechanism
\cite{21} then there should still be a break in slope in $T_{c_{+}}$ at the
pressure where magnetism is suppressed.\\
\hspace*{0.5cm} In the next paragraph we analyse the possibility of a rotation
and decrease of the magnetic moment suggested by Blount et al. \cite{16}
and Joynt \cite{17}. The rotation of magnetic moment can be equivalently
described by an additional linearly independent magnetic moment
$\bar{m}\;\;(\bar{m}\bot\bar{M})$ included.

\section{\bf Two magnetic moment model}
\hspace{0.5cm} In this section we consider the possibility that the
magnetic moment rotates at the temperature of the order of $T_{c}$
in such a way that the observed Bragg scattering intensity is reduced.
This requires two components of magnetisation. Therefore we propose
a revised G-L free energy density:
\begin{equation}
\label{e25}
F=F_{S}+F_{M}+F_{m}+F_{SM}+F_{sm}\;,
\end{equation}
where
\begin{equation}
\label{e026}
F_{M}=\left\{
\begin{array}{l}
\alpha_{M}\left(T^{\ast}-T_{N}\right)M^{2}+\frac{1}{2}\beta_{M}M^{4}\;\;
for\;\;T\leq T_{m}\\
\alpha_{M}\left(T-T_{N}\right)M^{2}+\frac{1}{2}\beta_{M}M^{4}\;\;
for\;\;T>T_{m}\;,
\end{array}
\right.
\end{equation}
\begin{equation}
\label{e26}
F_{m}=\alpha_{m}(T-T_{m})m^{2}+\frac{1}{2}\beta_{m}m^{4}\;,
\end{equation}
\begin{equation}
\label{e27}
F_{sm}=\gamma'|\bar{m}\bar{\psi}|^{2}+\alpha'm^{2}|\bar{\psi}|^{2}\;,
\end{equation}
and $F_{S},\;F_{M},\;F_{SM}$ are given by Eqs.$(\ref{e2})-(\ref{e4})$.
$T_{m}$ is the N\'eel temperature of the magnetic moment $\bar{m}$ and
$T_{m}\sim T_{c}$. The new coefficients $\alpha_{m}$ and $\beta_{m}$ in
\mbox{$(\ref{e26})$} are positive. This free energy is correct to the fourth
order in the space of $\bar{M},\;\bar{m}\;{\rm and}\;\bar{\psi}$. For the
sake of simplicity we have neglected the coupling term between the two
magnetic moments and the superconducting order parameter
\mbox{$(\;mM(\psi_{x}\psi_{y}^{\ast}+\psi_{x}^{\ast}\psi_{y})\;)$}
here, assuming it to have a little effect on the results. Another free
energy term involving $\bar{M}$ and $\bar{m}$
\mbox{$(\;\sim m^{2}M^{2}\;)$} is included implicitly in $T_{m}$ and
$T^{\ast}$ by a proper diagonalization of the magnetic part of the free
energy \mbox{( Appendix B ).} As it is seen from $(\ref{e25})$, a magnetic
moment $M$ is constant in the absence of superconductivity and equals:
\begin{equation}
\label{e28}
M^{2}=\frac{\alpha_{M}}{\beta_{M}}(T_{N}-T^{\ast})\;.
\end{equation}
This approximation is correct for temperatures lower than a certain
temperature of the order of $T_{c}$. We believe that this assumed
temperature dependence of $M^{2}$ is due to a change in a Fermi surface
and it is exclusively of the microscopic origin. However in the
\mbox{Appendix B} we present a phenomenological explanation of this fact,
when relation $(\ref{a4})$ is fulfilled. In this interpretation $M^{2}$
becomes constant below the temperature $T_{m}\;\;(\ref{e26},\ref{a7})$,
that is the temperature at which the magnetic moment $\bar{m}$ appears.
Although \mbox{$T_{m}\sim T_{c}$}, this reasoning is valid only if
\mbox{$T_{m}>T_{c_{+}}$} which seems to be in agreement with the
experimental data \cite{1,10}.\\
\hspace*{0.5cm} Proceeding in the same way as in section 2., from the
pressure requirements \mbox{$(\ref{e0},\;\ref{e14})$} and the free energy
density $(\ref{e25})$, we obtain the following conditions:
\begin{eqnarray}
\label{e29}
(|\gamma|-\alpha)M_{0}^{2} & > & \alpha'm_{0}^{2}\;,\\
\label{e30}
\left[\frac{1}{4}\left(3-\frac{\beta_{1}}{\beta_{2}}\right)|\gamma|
-\alpha\right]M_{0}^{2} & > & \left[\alpha'+\frac{\beta_{1}+\beta_{2}}
{4\beta_{2}}\gamma'\right]m_{0}^{2}\;,
\end{eqnarray}
where we have assumed, that $\bar{m}$ disappears at the same critical
pressure $p_{N}$ as $\bar{M}$ does $(\ref{e6})$:
\begin{equation}
\label{e31}
m^{2}=m_{0}^{2}\frac{p_{N}-p}{p_{N}}
\end{equation}
otherwise a kink in the pressure dependence of $T_{c_{-}}$ and $T_{c_{+}}$
should be observed, which is not the case (see Fig.1) \cite{2,7,8}.\\
\hspace*{0.5cm} Since there is no coupling terms between $\bar{m}$ and
$\bar{M}$ in the free energy density $(\ref{e25})$, it yields the same
temperature dependence of $M^{2}$ as in \mbox{Eqs.$(\ref{e16})-(\ref{e19})$.}
Therefore in order to obtain an appropriate temperature behavior of $M^{2}$
\cite{1,10} (Fig. 2a) either $(\ref{e23})$ or $(\ref{e24})$ must be
satisfied.\\
\hspace*{0.5cm} Now we are able to give the final conditions for the
\mbox{G-L coefficients} in the free energy density which agrees with the
experiments \cite{1,2,7,8,10} discussed in this paper. For $M^{2}$ increasing
with the temperature up to $T=T_{c_{-}}$ and decreasing above this temperature
the conditions $(\ref{e24})$ and \mbox{$(\ref{e29})-(\ref{e30})$} are to be
held. They lead to a simple constraint on $\alpha'$, which is necessary but
not sufficient:
\begin{eqnarray}
\label{e32}
|\gamma|M_{0}^{2} & > & 2\alpha' m_{0}^{2}\;.
\end{eqnarray}
When $M^{2}$ as a function of temperature has a kink at \mbox{$T=T_{c_{+}}$},
that is increases below this temperature and decreases above it, the
conditions $(\ref{e23})$ and \mbox{$(\ref{e29})-(\ref{e30})$} must be
fulfilled and they yield the negative value of $\alpha'$ :
\begin{eqnarray}
\label{e33}
\alpha' & < & 0 \;.
\end{eqnarray}

\section{\bf ${\bf \left(\frac{1}{2},0,1\right)}$ neutron scattering intensity}
\hspace{0.5cm} We are going to consider both the magnetic moments $\bar{M}$
and $\bar{m}$ more thoroughly now. Here again we restrict the calculations
to the linear in $\alpha$, $\beta$, $\alpha'$ and $\gamma'$ coupling
coefficients terms, what yields a negligible in this approximation
$M^{2}$ and $m^{2}$ discontinuity at $T_{c_{-}}$. A minimisation of the free
energy \mbox{$(\ref{e25})$} as a magnetic moment $\bar{m}$ function leads
to the temperature dependence of $m^{2}$ :
\begin{equation}
\label{e34}
m^{2}=m_{c}^{2}+(a_{m}-\frac{\alpha_{m}}{\beta_{m}})T\;,
\end{equation}
where
\begin{equation}
\label{e35}
m_{c}^{2}=\frac{\alpha_{m}}{\beta_{m}}T_{m}-a_{m}T_{c}\;,
\end{equation}
and
\begin{eqnarray}
\label{e36}
a_{m}=\frac{\alpha_{S}}{\beta_{m}}\frac{\alpha'}{\beta_{1}+\beta_{2}} & for &
T_{c_{-}}<T\leq T_{c_{+}}\;,\\
\label{e37}
a_{m}=\frac{\alpha_{S}}{\beta_{m}}\frac{2\alpha'+\gamma'}{2\beta_{1}} & for &
T\leq T_{c_{-}}\;.
\end{eqnarray}
We assume throughout this paper that the magnetic moments lie in the basal
plane
since the easy magnetic directions are confined to this plane. In the previous
chapters we were considering the neutron reflections at the
\mbox{reciprocal-lattice} point \mbox{$\bar{q_{1}}=(1,\frac{1}{2},0)$}
\cite{1,10} (Fig. 2a). The magnetic Bragg scattering measurements
revealed a different
temperature dependence of the neutron scattering intensity at
\mbox{$\bar{q_{2}}=(\frac{1}{2},0,1)$} \cite{18} (Fig. 2b).
Below the temperature of the
order of $T_{c}$ the \mbox{$(\frac{1}{2},0,1)$} intensity ceases to evolve
and becomes constant. Actually, Aeppli et al. \cite{18} did not go with
temperature low enough to be definitely positive about the $T$ independence of
the measured intensity in the whole temperature range below $T_{c}$.
Nevertheless, we assume here a constant value of \mbox{$(\frac{1}{2},0,1)$}
neutron scattering intensity below $T_{c_{+}}$, that is we suggest this
effect to be due to superconductivity. The neutron scattering intensity at the
\mbox{reciprocal-lattice} point $\bar{q}$ reflects the magnetic vectors
perpendicular to the $\bar{q}$ vector. For the sake of simplicity we choose
a magnetic moment
\begin{equation}
\label{e40}
\bar{M_{1}}=\bar{M}+\bar{m}\;
\end{equation}
perpendicular to \mbox{$\bar{q_{2}}=(\frac{1}{2},0,1)$} which means that
$M_{1}^{2}$ is detected in \mbox{$(\frac{1}{2},0,1)$} measurements.
On this particular magnetic orientation we want to check, without going
into the detailed calculation of a general case, whether the two magnetic
moment model can interpret both neutron scattering experiments.
It will yield some additional constraints on the \mbox{G-L} free energy
coefficients \mbox{$(\ref{e25})-(\ref{e27})$}.
One of the possible considered configurations of the magnetic and
reciprocal-lattice vectors, where instead of
\mbox{$\bar{q_{1}}=(1,\frac{1}{2},0)$} and
\mbox{$\bar{q_{2}}=(\frac{1}{2},0,1)$} their projections on the XY plane -
\mbox{$(1,\frac{1}{2})$} and \mbox{$(\frac{1}{2},0)$} were plotted,
is presented in Fig.3.
$\bar{M}$ is the magnetic moment seen in $(1,\frac{1}{2},0)$ neutron
scattering, while $\bar{M_{1}}$ is detected in $(\frac{1}{2},0,1)$
measurements. The temperature dependence of $M^{2}$ has been considered in the
previous paragraphs of this paper \mbox{$(\ref{e16}-\ref{e19},\ref{e28})$}.
According to \cite{18} $M_{1}^{2}$ is temperature independent for
\mbox{$T\leq T_{c} \sim T_{c_{+}}$}:
\begin{eqnarray}
\label{e38}
M_{1}^{2}=const & for & T<T_{c_{+}}\;.
\end{eqnarray}
Assuming the
temperature dependent corrections to $M$ $\mbox{$(\ref{e16})$}-
\mbox{$(\ref{e19})$}$ and
$m$ $\mbox{$(\ref{e34})$}-\mbox{$(\ref{e37})$}$ to be small,
experimentally estimated as
about 5\% of the total magnetic moments values \cite{1,10}, we linearize
$M$ and $m$ in $T$ and insert them into \mbox{Eq.$(\ref{e40})$}. Then the
condition $(\ref{e38})$ leads to the following constraints on the \mbox{G-L}
coefficients
\begin{eqnarray}
\label{e46}
f(\frac{\gamma+\alpha}{\beta_{1}+\beta_{2}},\frac{\alpha'}
{\beta_{1}+\beta_{2}})=0\;,\\
\label{e47}
f(\frac{\gamma+2\alpha}{2\beta_{1}},\frac{\gamma'+2\alpha'}
{2\beta_{1}})=0\;,
\end{eqnarray}
where
\begin{equation}
\label{e48}
f(x,y)=\frac{\alpha_{S}}{\beta_{M}}x+\frac{1}{\beta_{m}}
(\alpha_{S}y-\alpha_{m})\;.
\end{equation}
We solve the Eqs.$(\ref{e46})$ and $(\ref{e47})$ and \mbox{obtain :}
\begin{equation}
\label{e49}
\beta_{1}=\frac{\alpha_{S}}{2\alpha_{m}}\left[\frac{\beta_{m}}{\beta_{M}}
(\gamma+2\alpha)+\gamma'+2\alpha'\right]\;,
\end{equation}
\begin{equation}
\label{e50}
\beta_{2}=\frac{\alpha_{S}}{2\alpha_{m}}\left[\frac{\beta_{m}}{\beta_{M}}
\gamma-\gamma'\right]\;.
\end{equation}
According to experiments \cite{15}, $\beta_{1}$ and $\beta_{2}$ coefficients
should obey a following relation :
\begin{equation}
\label{e51}
\beta_{1}>\beta_{2}>0\;.
\end{equation}
{}From \mbox{$(\ref{e49})-(\ref{e51})$} we have then :
\begin{equation}
\label{e52}
\frac{\beta_{m}}{\beta_{M}}\gamma-\gamma'>0
\end{equation}
and
\begin{equation}
\label{e53}
\frac{\beta_{m}}{\beta_{M}}\alpha+\gamma'+\alpha'>0\;.
\end{equation}
Since $\gamma\;(\ref{e4})$ is negative, inequality $(\ref{e52})$ leads to a
negative $\gamma'$ value and finally relation $(\ref{e52})$ is equivalent to :
\begin{equation}
\label{e54}
\gamma'=-|\gamma'|\;,\;\;\;\;|\gamma'|>\frac{\beta_{m}}{\beta_{M}}|\gamma|\;.
\end{equation}
Therefore we have obtained conditions \mbox{$(\ref{e49})-(\ref{e50})$} and
\mbox{$(\ref{e53})-(\ref{e54})$} which are to be fulfilled by \mbox{G-L}
free energy coefficients. However we cannot forget about the constraints
which follow from the $M^{2}$ temperature evolution requirements
\mbox{$(\ref{e23})-(\ref{e24})$} and these which are necessary to fit the
pressure data \mbox{$(\ref{e29})-(\ref{e30})$}. One can check easily that
the conditions $(\ref{e23})$ ( kink in $M^{2}$ at $T=T_{c_{+}}$ ) and
\mbox{$(\ref{e53})-(\ref{e54})$} lead to a negative value of $\alpha'$,
while the constraint $(\ref{e24})$ ( kink in $M^{2}$ at \mbox{$T=T_{c_{-}}$ )}
along with \mbox{Eqs.$(\ref{e53})-(\ref{e54})$} yield a positive $\alpha'$
value. From \mbox{Eqs.$(\ref{e29})-(\ref{e30})$} we get more information
about the magnetic moments values at pressure $p=0$, that is $M_{0}\;
(\ref{e6})$ and $m_{0}\;(\ref{e31})$. It is more convenient for this purpose
to use the experimentally established $\frac{\beta_{2}}{\beta_{1}}$ ratio :
\mbox{$\frac{\beta_{2}}{\beta_{1}}\simeq 0.4$} \cite{15}, just to get rid
of $\beta_{1}$ and $\beta_{2}$ coefficients in $(\ref{e30})$. The relation
\mbox{$\frac{\beta_{2}}{\beta_{1}}=0.4$} along with \mbox{$\beta_{1}\;
(\ref{e49})$} and \mbox{$\beta_{2}\;(\ref{e50})$} formulae allow the
reduction of one of the coupling coefficients through the equation :
\begin{equation}
\label{e55}
\alpha'=\frac{7}{4}(|\gamma'|-\frac{\beta_{m}}{\beta_{M}}|\gamma|)
+\frac{\beta_{m}}{\beta_{M}}(|\gamma|-\alpha)\;,
\end{equation}
so we can consider $\gamma',\;\gamma$ and $\alpha$ parameters as the only
independent in all the conditions. It is straightforward to show that
$\alpha'$ given by \mbox{Eq.$(\ref{e55})$} obeys the
\mbox{Eqs.$(\ref{e23})-(\ref{e24})$} and \mbox{$(\ref{e53})-(\ref{e54})$}.
Returning to $M_{0}$ and $m_{0}$ magnitudes, for \mbox{$\alpha>|\gamma|\;
(\ref{e23})$}, we obtain from \mbox{$(\ref{e29})-(\ref{e30})$}, that
\begin{equation}
\label{e56}
m_{0}^{2}>g_{0}M_{0}^{2}\;,
\end{equation}
where
\begin{displaymath}
g_{0}=max\left\{\frac{\alpha-|\gamma|}{|\alpha'|},
\frac{\alpha-\frac{1}{8}|\gamma|}{|\alpha'|+\frac{7}{8}|\gamma'|}\right\}\;.
\end{displaymath}
The condition above should be fulfilled when a kink in
\mbox{$\left(1,\frac{1}{2},0\right)$} neutron scattering intensity appeares at
$T_{c_{+}}\;(\ref{e23})$. In order to have $m_{0},M_{0}$ solutions of
\mbox{$(\ref{e29})-(\ref{e30})$} when condition $(\ref{e24})$ is held, that
is in a case of the \mbox{$\left(1,\frac{1}{2},0\right)$} neutron scattering
peak at $T_{c_{-}}$, another constraint is to be fulfilled :
\begin{equation}
\label{e57}
\alpha'<\frac{7}{8}|\gamma'|\;.
\end{equation}
Inequality $(\ref{e57})$ is a necessary condition to make sense to
the relations $(\ref{e29})$ and $(\ref{e30})$.\\
Finally, we obtain from \mbox{$(\ref{e29})-(\ref{e30})$} the constraint on the
relative $m_{0}$ and $M_{0}$ values :
\begin{equation}
\label{e58}
\frac{\alpha-\frac{1}{8}|\gamma|}{\frac{7}{8}|\gamma'|-\alpha'}<
\frac{m_{0}^{2}}{M_{0}^{2}}<\frac{|\gamma|-\alpha}{\alpha'}\;,
\end{equation}
and another condition which follows straightly from $(\ref{e58})$ :
\begin{equation}
\label{e59}
|\gamma\gamma'|-|\gamma|\alpha'-|\gamma'|\alpha>0\;.
\end{equation}
We have been looking here at the additional constraints on the fourth order
coefficients in the \mbox{Ginzburg-Landau} free energy, that follow from the
requirement of a constant magnetic moment detected in
\mbox{$\left(\frac{1}{2},0,1\right)$} neutron scattering measurements
\cite{18} (Fig. 2b). We have assumed $T_{c_{+}}$ as a characteristic
temperature at
which the magnetic moment $M_{1}\;(\ref{e38})$ becomes constant. Nevertheless
it is straightforward to show that $M_{1}$ cannot be constant above
$T_{c_{+}}$. Let us look at the temperatures \mbox{$T>T_{m}$} first. Since
$T_{m}$ is the N\'eel temperature for $\bar{m}\;(\ref{a6})$, there is only one
magnetic moment $\bar{M}$ left at $T>T_{m}$. $M_{1}$ is simply $M$'s
projection on a particular direction ( Fig.3 ) and shows the same temperature
dependence as $M$ does $(\ref{a9})$. Therefore $M_{1}$ is a decreasing
function of temperature for \mbox{$T>T_{m}$} as $M$ is $(\ref{a9})$. In the
temperature range \mbox{$T_{c_{+}}<T<T_{m}$}, on the other hand, we obtain
from the free energy $(\ref{a8})$ a constant $M^{2}$ value $(\ref{e28})$ and
\mbox{$m^{2}=\frac{\alpha_{m}}{\beta_{m}}\left(T_{m}-T\right)$}. Therefore
$(\ref{e40})$ cannot lead to a constant $M_{1}$ value,
otherwise $\alpha_{m}=0$ and $\bar{m}$ vanishes, what makes no sense for this
approach.

\section{\bf Conclusions}
\hspace{0.5cm} We have considered superconducting \mbox{$U\!Pt_{3}$}
in zero magnetic field. Our interest has been focused on the hydrodynamic
pressure \cite{2,7,8} and neutron scattering experiments \cite{1,10,18}.
We have shown that the pressure dependence of the transition temperatures
and the abrupt change in the \mbox{$\left(1,\frac{1}{2},0\right)$} neutron
scattering intensity
at \mbox{$T\sim T_{c}$} \cite{1,10} cannot be explained
quantitatively within a simple two component superconducting order
parameter which couples to one component antiferromagnetism. As one
way of reconciling this problem we have suggested the existence of another
magnetic moment which emerges at \mbox{$T\sim T_{c}$}. This generalized
approach of the two independent magnetic moments coupling to the
superconductivity allowed us to obtain a concise picture of discussed
phenomena and yields several stringent constraints on the fourth order
coefficients in the \mbox{Ginzburg-Landau} free energy density $(\ref{e25})$.
We have concluded that the kink in a \mbox{$\left(1,\frac{1}{2},0\right)$}
neutron scattering intensity may exist at $T_{c_{+}}$ when $(\ref{e23})$ and
\mbox{$(\ref{e29})-(\ref{e30})$} relations between the \mbox{G-L} coefficients
are obeyed or at $T_{c_{-}}$ under the condition of $(\ref{e24})$ and
\mbox{$(\ref{e29})-(\ref{e30})$}. If we interpret the results of
\mbox{$\left(\frac{1}{2},0,1\right)$} Bragg magnetic scattering experiments
\cite{18} as characteristic feature for all temperatures below $T_{c}$
and assume the magnetic moments orientation as in Fig. 3, we can express
$\beta_{1}$ and $\beta_{2}$ \mbox{G-L} coefficients in terms of the coupling
constants \mbox{$(\ref{e49})-(\ref{e50})$}. The requirement
\mbox{$\beta_{1}>\beta_{2}>0$} leads to a negative value of a coupling
constant $\gamma'\;(\ref{e27},\ref{e54})$ and negative $\alpha'\;
(\ref{e27})$ coefficient value when a peak in
\mbox{$\left(1,\frac{1}{2},0\right)$} neutron scattering intensity is at
$T_{c_{+}}$ or positive $\alpha'$ value for a peak at $T_{c_{-}}$. These
considerations yield also some constraints on the zero pressure magnetic
moments values \mbox{$(\ref{e56},\ref{e58})$} and coupling coefficients
\mbox{$(\ref{e57},\ref{e59})$}. We have evaluated
\mbox{$(\ref{e56})-(\ref{e59})$} constraints for the experimentally
established ratio \mbox{$\frac{\beta_{2}}{\beta_{1}}\simeq 0.4$} \cite{15}.
This given value of $\frac{\beta_{2}}{\beta_{1}}$ allows us to express one
of the G~-~L coupling coefficients in terms of the others $(\ref{e55})$.\\
We have considered two magnetic moments in a crystal basal plane only.
However, we cannot exclude any of them out of this plane.  There is always
a possibility of a magnetic structure following a recently discovered
structural modulation in a crystal \cite{19}. Unfortunately the resolution
of a neutron scattering measurements may be to small to be decisive.
For the completness of the picture it should be added that despite a large
number of experimental evidence the main facts seems to be unsettled.
It concernes the phase diagram in the p-T plane measured by Boukhny et al.
\cite{20} where the slope of $T_{c_{-}}$ curve is positive and the condition
$(\ref{e0})$ does not hold. Moreover the recent x-ray resonant magnetic and
neutron magnetic scattering measurements \cite{21} show no correlation between
the split superconducting transition and the weak antiferromagnetic order in
$U\!Pt_{3}$ and as they also find no evidence of magnetic moment rotation
their results together with the conclusions of the paper
suggest other possible issues like symmetry-breaking fields
of structural origin \cite{19} or the existence of two one dimensional
superconducting states.

\section*{\bf Acknowledgements}
We would like to thank G. Aeppli for fruitful comments and providing us with
his recent experimental results as well as R. Joynt and L. Taillefer for
helpful letters.\\
This work was supported by the European Community's Action for Cooperation
in Sciences and Technology with Central and Eastern European Countries
Fellowship \mbox{ contract $n^{\circ}:\;ERB-CIPA-92-2218$.}

\section*{\bf Appendix A}
\begin{equation}
\label{a01}
T_{c_{+}}=T_{c}-\frac{\alpha+\gamma}{\alpha_{S}}\frac{\alpha_{M}}
{\beta_{M}}\left(T_{N}-T^{\ast}\right)\;,
\end{equation}
\begin{equation}
\label{a02}
T_{c_{-}}=T_{c}+\frac{\alpha_{M}}{\alpha_{S}}
\frac{\gamma\left(\beta_{1}-\beta_{2}\right)-2\beta_{2}\alpha}
{2\beta_{2}\beta_{M}-\gamma\left(\alpha+\gamma\right)}\left(T_{N}-T^{\ast}
\right)\;,
\end{equation}
\begin{equation}
\label{a03}
M^{2}=M_{c}^{2}+a_{M}T\;,
\end{equation}
\begin{equation}
\label{a04}
M_{c}^{2}=\left\{
\begin{array}{lll}
\frac{\beta_{1}+\beta_{2}}{\lambda_{+}}
\left[\alpha_{M}\left(T_{N}-T^{\ast}\right)-\frac{\alpha_{S}}
{\beta_{1}+\beta_{2}}\left(\alpha+\gamma\right)T_{c}\right] & for &
T_{c_{-}}<T\leq T_{c_{+}}\\
\frac{2\beta_{2}}{\lambda_{-}}\left[2\alpha_{M}\beta_{1}
\left(T_{N}-T^{\ast}\right)-\alpha_{S}\left(2\alpha+\gamma\right)T_{c}\right]
& for & T\leq T_{c_{-}}\;,
\end{array}
\right.
\end{equation}
\begin{equation}
\label{a05}
a_{M}=\left\{
\begin{array}{lll}
\frac{\alpha_{S}}{\lambda_{+}}\left(\alpha+\gamma\right) & for &
T_{c_{-}}<T\leq T_{c_{+}}\\
\frac{2\alpha_{S}}{\lambda_{-}}\beta_{2}\left(2\alpha+\gamma\right) & for &
T\leq T_{c_{-}}\;,
\end{array}
\right.
\end{equation}
\begin{equation}
\label{a06}
\lambda_{+}=\beta_{M}\left(\beta_{1}+\beta_{2}\right)-
\left(\alpha+\gamma\right)^{2}\;,
\end{equation}
\begin{equation}
\label{a07}
\lambda_{-}=4\beta_{1}\beta_{2}\beta_{M}-4\beta_{2}\alpha
\left(\alpha+\gamma\right)-\gamma^{2}\left(\beta_{1}+\beta_{2}\right)\;.
\end{equation}

\section*{\bf Appendix B}
\hspace{0.5cm}The complete magnetic free energy for the magnetic moments
$\bar{M}$ and $\bar{m}$ \mbox{$(\bar{m}\perp\bar{M})$} at the temperatures
\mbox{$T<T^{\ast}$} is :
\begin{equation}
\label{a1}
F_{magn}=A_{M}\left(T-T_{N}\right)M^{2}+\frac{1}{2}B_{M}M^{4}+
A_{m}\left(T-T^{\ast}\right)m^{2}+\frac{1}{2}B_{m}m^{4}+Cm^{2}M^{2}\;,
\end{equation}
where $T_{N}$ and $T^{\ast}$ are the N\'eel temperatures for $\bar{M}$ and
$\bar{m}$ magnetic moments appropriately.\\
We assume $T_{N}>T^{\ast}$. From the minimisation of $F_{magn}$ one gets :
\begin{equation}
\label{a2}
M^{2}=\frac{B_{M}B_{m}}{B_{M}B_{m}-C^{2}}\left[\frac{A_{M}}{B_{M}}T_{N}
-\frac{A_{m}}{B_{m}}\frac{C}{B_{M}}T^{\ast}-\left(\frac{A_{M}}{B_{M}}-
\frac{A_{m}}{B_{m}}\frac{C}{B_{M}}\right)T\right]\;,
\end{equation}
and
\begin{equation}
\label{a3}
m^{2}=\frac{1}{B_{M}B_{m}-C^{2}}\left(A_{m}B_{M}T^{\ast}-A_{M}CT_{N}\right)
-\frac{A_{m}}{B_{m}}T\;.
\end{equation}
For a particular choice of the coupling coefficient
\begin{equation}
\label{a4}
C=\frac{A_{M}}{A_{m}}B_{m}\;,
\end{equation}
$M^{2}$ attains a constant value :
\begin{equation}
\label{a5}
M^{2}=\frac{A_{M}A_{m}^{2}}{A_{m}^{2}B_{M}-A_{M}^{2}B_{m}}\left(T_{N}-T^{\ast}
\right)\;,
\end{equation}
and
\begin{equation}
\label{a6}
m^{2}=\frac{A_{m}}{B_{m}}\left(T_{m}-T\right)\;,
\end{equation}
where
\begin{equation}
\label{a7}
T_{m}=\frac{A_{m}^{2}B_{M}-A_{M}^{2}B_{m}\frac{T_{N}}{T^{\ast}}}
{A_{m}^{2}B_{M}-A_{M}^{2}B_{m}}T^{\ast}\;.
\end{equation}
The temperature $T^{\ast}$ should be of the order of $T_{N}$ to give a
positive value of $T_{m}$. From $(\ref{a5})$ and $(\ref{a6})$ we can see
that the magnetic free energy can be written as :
\begin{equation}
\label{a8}
F_{magn}=\alpha_{M}\left(T^{\ast}-T_{N}\right)M^{2}+\frac{1}{2}\beta_{M}M^{4}
+\alpha_{m}\left(T-T_{m}\right)m^{2}+\frac{1}{2}\beta_{m}m^{4}\;,
\end{equation}
for $T<T_{m}$, and
\begin{equation}
\label{a9}
F_{magn}=\alpha_{M}\left(T-T_{N}\right)M^{2}+\frac{1}{2}\beta_{M}M^{4}\;,
\end{equation}
for $T>T_{m}$.\\
This is the free energy of two magnetic moments $(\ref{e026})-(\ref{e26})$
we use in this paper. The new G-L coefficients are given by the old
\mbox{ones :}
\begin{eqnarray}
\label{a10}
\alpha_{M}=A_{M}A_{m}^{2}\;,\\
\label{a11}
\beta_{M}=A_{m}^{2}B_{M}-A_{M}^{2}B_{m}\;,\\
\label{a12}
\alpha_{m}=A_{m}\;,\\
\label{a13}
\beta_{m}=B_{m}\;.
\end{eqnarray}
These considerations are relevant only when
$(\ref{a4})$ condition is fulfilled.

\newpage
\section*{\bf Figure captions}
Fig.1 Pressure dependence of the superconducting phase transition
temperature \cite{8}.\\
Fig.2 Field and temperature dependence of \mbox{$(1,\frac{1}{2},0)$}
\cite{10} (a) and \mbox{$(\frac{1}{2},0,1)$} \cite{18} (b)
neutron scattering intensities\\
Fig.3 The relative orientation of the magnetic moments $\bar{M}$,
$\bar{m}$ and $\bar{M_{1}}$ and the neutron scattering vectors
$q_{1}$ and  $q_{2}$, where \mbox{$\tan\varphi=\frac{1}{2}$}.

\newpage
\setlength{\unitlength}{1mm}
\begin{picture}(150,150)
\put(50,70){\vector(2,1){25}}
\put(50,70){\vector(1,0){20}}
\put(50,70){\vector(0,1){30}}
\put(50,70){\vector(-1,2){12}}
\put(80,65){$\bar{q_{2}}=\left(\frac{1}{2},0\right)$}
\put(50,105){$\bar{M_{1}}$}
\put(80,85){$\bar{q_{1}}=\left(1,\frac{1}{2}\right)$}
\put(35,100){$\bar{M}$}
\put(57.3,71.5){$\varphi$}
\put(47.5,80){$\varphi$}
\put(50,10){\vector(0,1){30}}
\put(50,10){\vector(1,0){20}}
\put(50,10){\vector(2,3){20}}
\put(50,45){$\bar{M}$}
\put(70,5){$\bar{m}$}
\put(70,45){$\bar{M_{1}}$}
\end{picture}
\begin{center}
Fig.3\\
\end{center}

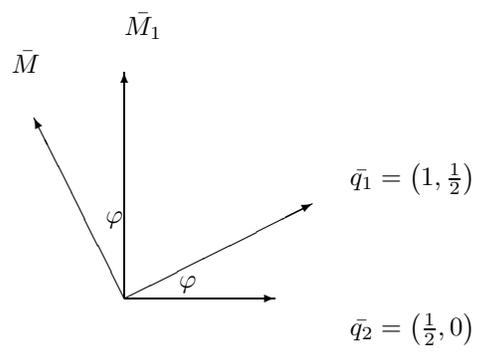
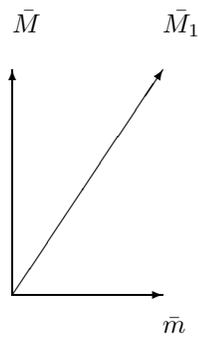
\end{document}